\documentclass{PoS}
\pdfoutput=1
\usepackage{subfigure}
\usepackage{latexsym}
\usepackage{amsfonts}
\usepackage[latin1]{inputenc}
\usepackage{graphicx}
\usepackage{mathrsfs}
\usepackage{amssymb}
\usepackage{amsmath}
\usepackage{verbatim}
\usepackage{multirow}
\usepackage{soul}
\newcommand{\pb}{{\ensuremath\rm pb}}
\newcommand{\GeV}{{\ensuremath\rm GeV}}
\newcommand{\TeV}{{\ensuremath\rm TeV}}
\newcommand{\al}{\alpha}
\newcommand{\be}{\beta}
\newcommand{\lam}{\lambda}

\newcommand{\HS}{\texttt{HiggsSignals}}
\newcommand{\HB}{\texttt{HiggsBounds}}

\newcommand{\pstar}{\ensuremath{p_*}}

\newcommand{\lb}{\left (}
\newcommand{\rb}{\right )}

\title{The Higgs singlet extension at LHC Run 2}

\ShortTitle{The Higgs singlet extension at LHC Run 2}

\author{{G. Chalons}\\
LPSC, Universit\'e Grenoble-Alpes, CNRS/IN2P3\\
LPT, CNRS, Universit\'e Paris-Sud, Universit\'e Paris-Saclay\\
        E-mail: \email{chalons@lpsc.in2p3.fr}}
\author{D. L\'opez-Val\\
KIT Karlsruhe\\
        E-mail: \email{david.val@kit.edu}}
\author{\speaker{T. Robens}\\
IKTP, TU Dresden\\
        E-mail: \email{Tania.Robens@tu-dresden.de}}
\author{T. Stefaniak\\
Santa Cruz Institute for Particle Physics and Department of Physics,
University of California, Santa Cruz, CA 95064, USA\\
E-mail: \email{tistefan@ucsc.edu}}

\abstract{
We discuss the current status of theoretical and experimental constraints on the real Higgs singlet extension of the Standard Model. For the second neutral (non-standard) Higgs
boson the full mass range from 1 GeV to 1 TeV accessible at past and current collider experiments is considered. We present benchmark scenarios for searches for an additional Higgs
state in the real Higgs singlet extension of the Standard Model in Run 2 of the LHC. We furthermore discuss electroweak corrections to the $H\,\rightarrow\,h\,h$ partial decay width within
this model.\\

\hfill KA-TP-37-2016, LPSC16258, LPT-Orsay-16-76
}

\FullConference{38th International Conference on High Energy Physics\\
		3-10 August 2016\\
		Chicago, USA}

\begin{document}
\section{The model and dominant constraints on the parameter space}

The simplest extension of the Standard Model (SM) Higgs sector, where an additional real {scalar} field is added \cite{Schabinger:2005ei, Patt:2006fw, Bowen:2007ia}, is a widely explored benchmark scenario for experimental searches at the LHC \cite{deFlorian:2016spz}. The model contains a complex $SU(2)_L$ doublet, denoted by $\Phi$, and a real scalar $S$ which is a singlet under the SM gauge group. The most general renormalizable Lagrangian compatible with an additional $Z_2$ symmetry contains the scalar potential
%\begin{eqnarray}\label{potential}
$V(\Phi,S ) \,=\, -m^2 \Phi^{\dagger} \Phi -\mu ^2 S ^2 + \lambda_1
(\Phi^{\dagger} \Phi)^2 + \lambda_2  S^4 + \lambda_3 \Phi^{\dagger}
\Phi S ^2.$
%\end{eqnarray}
In the unitary gauge, the Higgs fields are given by
%\begin{equation}\label{unit_higgs}
$\Phi \equiv
\left(
%\begin{gathered}
0 \;\;
\tfrac{\tilde{h}+v}{\sqrt{2}}
%\end{gathered} 
\right)^T, \,
S \equiv \frac{h'+v_s}{\sqrt{2}}$, with $v,\,v_s$ denoting the non-zero vacuum expectation values of the doublet and singlet, respectively.
%\end{equation} 
 The above potential leads to mixing between the gauge eigenstates via the mixing angle $\al$, such that $h\,=\,c_\al\,\tilde{h}-s_\al\,h',\,H\,=\,s_\al\,\tilde{h}+c_\al\,h'$, with $s_\al\,(c_\al)\,\equiv\,\sin\al \lb\cos\al \rb$.
We here use the convention that $m_h\,\leq\,m_H$, and choose {as input
parameters}
%\begin{\eqn}\label{eq:pars}
$m_h,\,m_H,\,\sin\alpha,\,v,\,\tan\beta\,\equiv\,\frac{v}{v_s}$,
%\end{\eqn}
where $v\,\sim\,246\,\GeV$. One of the scalar masses is fixed to $\sim\,125\,\GeV$.
The above mixing induces a rescaling of the {SM-like Higgs} couplings at tree level by $\sin\al \,\lb \cos\al \rb$ for $h~(H)$, with respect to the couplings for a SM Higgs boson of that mass. Furthermore, it features a genuinely new decay mode whenever the channel $H\,\rightarrow\,h\,h$ opens up kinematically.  See e.g. \cite{Pruna:2013bma,Robens:2015gla,Robens:2016xkb} for further details. %Thus, the branching ratios for decays into SM final states are given by the predictions of the SM Higgs of the respective mass at tree level. If the additional decay channel $H\,\rightarrow\,h\,h$ is allowed, the absolute branching ratios for decays into SM final states are accordingly modified. See e.g. \cite{Pruna:2013bma,Robens:2015gla,Robens:2016xkb} for details.

\begin{figure}
 \includegraphics[width=0.8\textwidth]{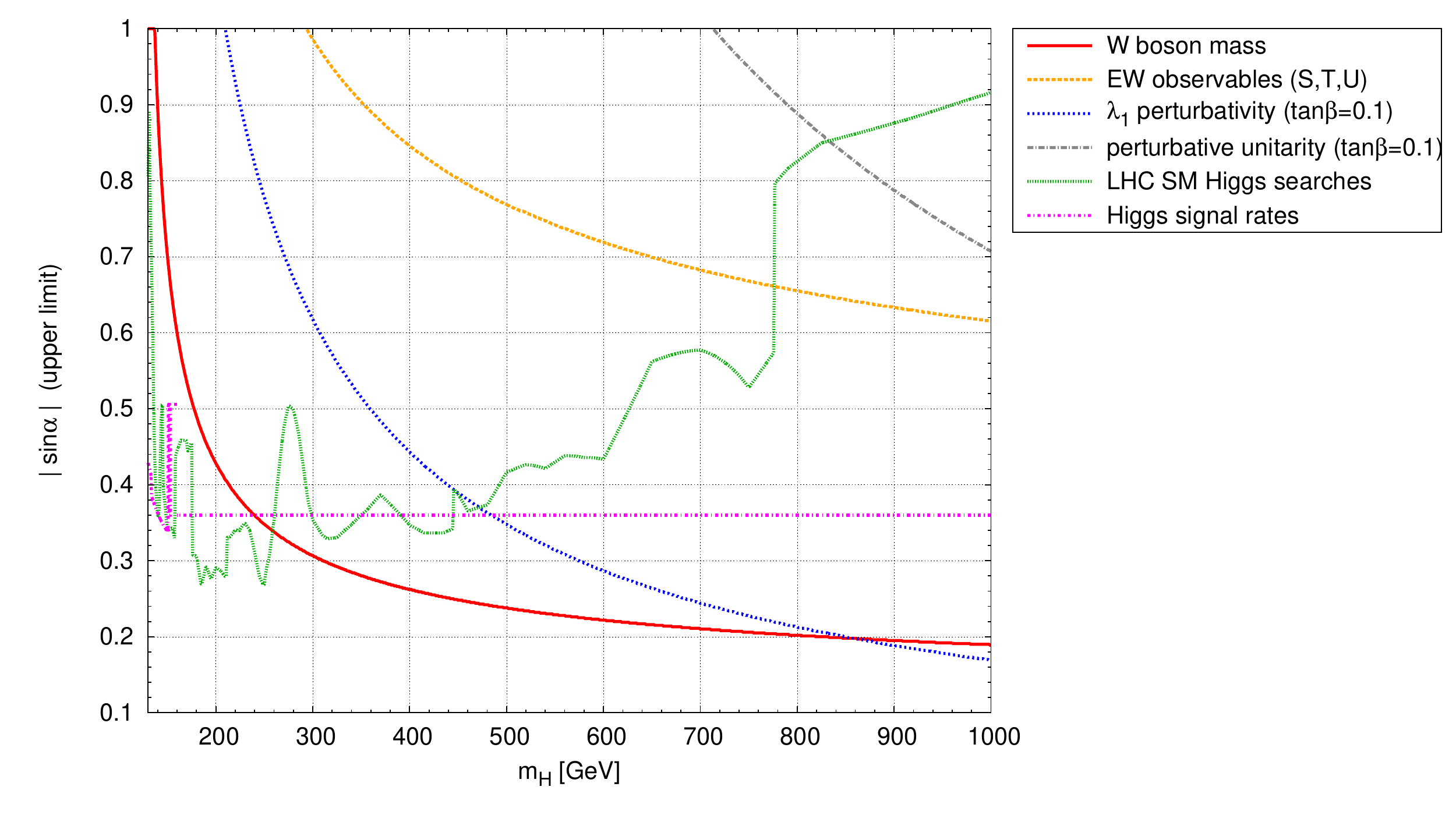}
\caption{\label{fig:sinamw} Maximal allowed values for $| \sin\al |$ in the high mass region, $m_H\in [130, 1000]\,\GeV$, from precision calculations of the $W$-boson mass (\emph{red, solid})~\cite{Lopez-Val:2014jva}, electroweak precision observables (EWPOs) tested via the oblique parameters $S$, $T$ and $U$ (\emph{orange, dashed}), perturbativity of the RG-evolved coupling $\lam_1$ (\emph{blue, dotted}), evaluated {for an exemplary choice} $\tan\be\,=\,0.1$, perturbative unitarity (\emph{grey, dash-dotted}), direct LHC Higgs searches (\emph{green, dashed}), and the Higgs signal strength (\emph{magenta, dash-dotted}). Taken from \cite{Robens:2016xkb}. More recent collider results (see e.g. \cite{CMS:2016jpd,CMS:2016knm}) are not included and can potentially influence the region where $m_H\,\lesssim\,400\,\GeV$.}
\end{figure}

Both theoretical and experimental constraints determine viable regions of the models parameter space, cf. \cite{Pruna:2013bma,Robens:2015gla,Robens:2016xkb}. Limits on the mixing angle for cases where the second scalar is heavier than $125\,\GeV$ mainly result from {\sl (i)} direct search limits, which we implemented using \HB\ (version 4.3.1)~\cite{Bechtle:2008jh,Bechtle:2011sb,Bechtle:2013wla}, {\sl (ii)} Higgs signal strength measurements, either implemented via \HS\ (version 1.4.0)~\cite{Bechtle:2013xfa} or taken as a direct limit from the combined Higgs signal strength measurement \cite{ATLAS-CONF-2015-044}, {\sl (iii)} the precision calculation of the $W$-boson mass within this model \cite{Lopez-Val:2014jva}, as well as {\sl (iv)} limits from perturbativity of the couplings. A summary of all constraints is given in Fig.~\ref{fig:sinamw}.
{Production} cross-sections for the 14 \TeV~ LHC, after all constraints have been taken into account, can reach up to $10\,\pb~$ for the total rate of all SM final states and up to 0.5 \pb~for $hh$ final states. Specific benchmarks for all mass ranges allowed by the limits in Fig.~\ref{fig:sinamw} have been presented in \cite{Robens:2016xkb,deFlorian:2016spz}.

\section{Renormalization}
\begin{figure}
 \centering
 \subfigure[~$m_h\,=\,125.09\GeV$]{
 \includegraphics[width=0.33\textwidth]{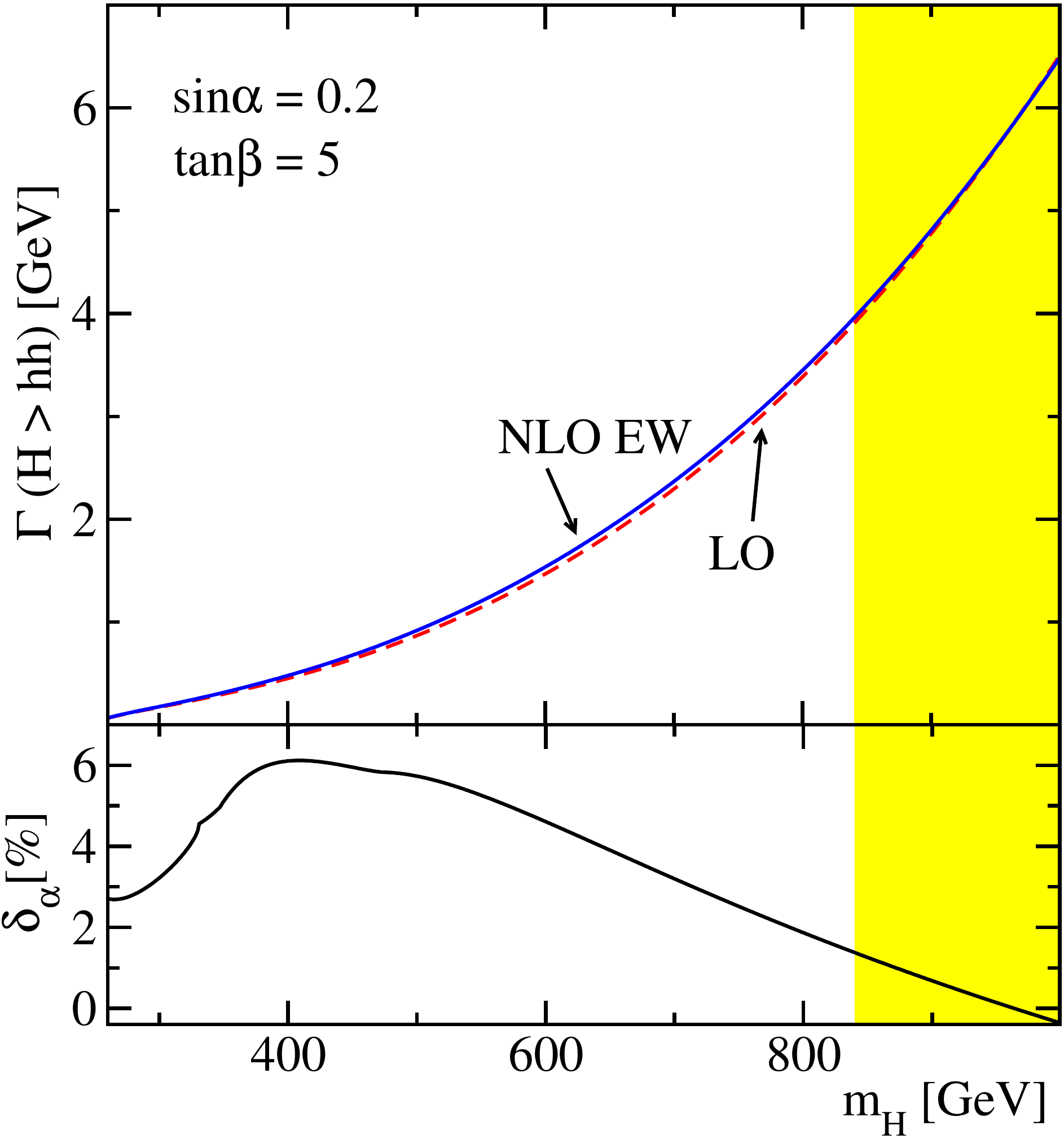}}
 \hfill
 \subfigure[~$m_H\,=\,125.09\GeV$]{
 \includegraphics[width=0.36\textwidth]{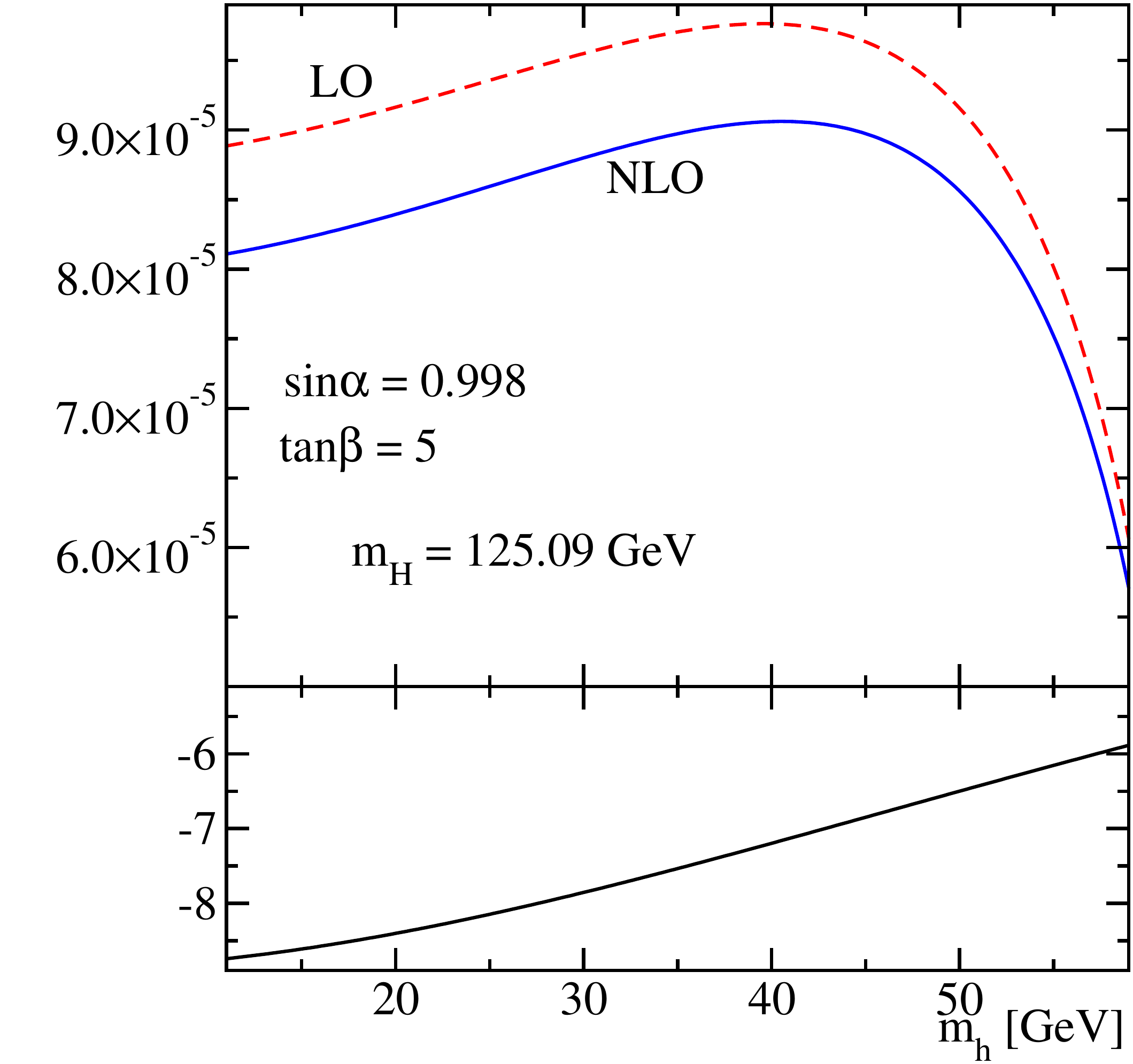}}
\caption{\label{fig:nlo} NLO corrections to the $H\,\rightarrow\,h\,h$ partial decay width, for fixed $\sin\al,\,\tan\be$ values and $m_h$ {\sl (left)} or $m_H$ {\sl (right)} being the 125 \GeV~resonance measured
at the LHC, as a function of the second scalar mass. We display the total decay width for $H\,\rightarrow\,h\,h$, we display the total decay width, along
with its relative one-loop correction. The yellow region is excluded {by} perturbativity. {\sl Note:} $\tan\be$ is defined as $\frac{v_s}{v}$ in this case, in contrast to the definitions given above. Taken from \cite{Bojarski:2015kra}.}
\end{figure}

The complete electroweak renormalization of the model has been discussed in \cite{Bojarski:2015kra}, where we applied a non-linear gauge fixing prescription implemented within the 
{\sc Sloops} framework (see e.g. 
\cite{Boudjema:2005hb,Baro:2009gn}), to study gauge-parameter dependence of several schemes. We found that  an \emph{improved} On-shell scheme, specified by the off-diagonal mass counterterm
%\item{}
%\begin{equation}
%\label{eq:improvedOS}
 $\delta m^2_{hH} = 
\mbox{Re}\,\Sigma_{hH}(\pstar^2)\big|_{\xi_{W}=\xi_Z=1,\tilde\delta_i=0}$ 
with $\pstar^2 = \frac{m_h^2+m_H^2}{2}\, ,$
%\end{equation}
exhibits the cleanest theoretical and numerical properties. We apply it to compute the one-loop electroweak corrections
to the decay width $\Gamma_{H\,\rightarrow\,h\,h}$.
 Once all present constraints on the model are included, we find mild
NLO corrections, typically of few percent, with theoretical uncertainties on the per mille level. Sample {results} are displayed in Fig. \ref{fig:nlo}.

\section*{Acknowledgements}
\vspace{-3mm}
{\small Funding is acknowledged from the F.R.S.-FNRS "Fonds de la Recherche Scientifique", the Theory- LHC-France initiative
of CNRS/IN2P3, the ERC advanced grant "Higgs@LHC",  the U.S.~Department of Energy grant number DE-SC0010107 and the Alexander von Humboldt foundation.}

%\bibliography{main,hnlo}
\providecommand{\href}[2]{#2}\begingroup\raggedright\endgroup

\end{document}